\documentstyle[11pt,aaspp4]{article}
 
\def\etal{{\rm et~al. }}
\def\mpc {h^{-1}{\rm{Mpc}}} 
\def\and  {\it {et al.} \rm}

\def\hmpc{\;h^{-1}{\rm Mpc}}

\def\spose#1{\hbox to 0pt{#1\hss}}
\def\simlt{\mathrel{\spose{\lower 3pt\hbox{$\mathchar"218$}}
     \raise 2.0pt\hbox{$\mathchar"13C$}}}
\def\simgt{\mathrel{\spose{\lower 3pt\hbox{$\mathchar"218$}}
     \raise 2.0pt\hbox{$\mathchar"13E$}}}

\newcommand{\beq}{\begin{equation}}
\newcommand{\eeq}{\end{equation}}
\newcommand{\beqa}{\begin{eqnarray}}
\newcommand{\eeqa}{\end{eqnarray}}

\begin{document}

\title
{The Space Density of Galaxy Peaks and the Linear Matter Power Spectrum.}

\author{Rupert A.C. Croft\altaffilmark{1}}
\affil{Department of Astronomy, The Ohio State University, Columbus,
Ohio 43210, USA}
\author{Enrique Gazta\~{n}aga\altaffilmark{2}}
\affil{CSIC, Institut d'Estudis Espacials de Catalunya (IEEC), 
Edifici Nexus-104 - c/ Gran Capitan 2-4, 08034 Barcelona, Spain}
\altaffiltext{1}{CSIC, Institut d'Estudis Espacials de Catalunya (IEEC), 
Edifici Nexus-104 - c/ Gran Capitan 2-4, 08034 Barcelona, Spain}
\altaffiltext{2}{Oxford University, Astrophysics, Keble Road, Oxford OX1 3RH,
 UK}

\begin{abstract}
One way of recovering information about the initial 
conditions of the Universe 
is by measuring features  of the cosmological density field which 
are preserved during gravitational evolution and galaxy formation.
In this paper we study  the total number density of 
peaks in a (galaxy) point distribution smoothed with a filter,
evaluating its usefulness as a means of inferring the shape 
of the initial (matter) power spectrum.
 We find that in numerical simulations 
which start from Gaussian initial conditions,
the peak density follows well that predicted by the theory of 
Gaussian density fields, even on scales where the clustering is
 mildly non-linear. For smaller filter
scales, $r \simlt 4-6 \hmpc$, we see evidence of merging as the peak density
decreases with time.
On larger scales,  the peak density is independent of time. 
One might also expect it to be fairly robust with respect 
to variations in biasing, i.e. the way galaxies trace mass fluctuations. 
We find that this is the case when we apply various 
biasing prescriptions to the matter
distribution in simulations. 
If the initial conditions are Gaussian, it is possible
to use the peak density measured from the
evolved field to reconstruct the shape of the {\it initial} 
power spectrum.  We describe a stable method for doing this and 
apply it to several biased and unbiased
non-linear simulations. We are able to recover the slope of the linear
matter power spectrum on scales $k \simlt 0.4 \hmpc^{-1}$. 
 The reconstruction has the advantage of being
independent of the cosmological
parameters ($\Omega$, $\Lambda$, $H_0$) and of the clustering normalisation
($\sigma_8$). The peak density and reconstructed
power spectrum slope therefore promise to be powerful
discriminators between popular cosmological scenarios.

\end{abstract}


\section{Introduction}

Attempts to reconstruct the initial conditions from which large-scale 
structure in the Universe grew have been carried out using several different
methods. One avenue of research involves attempting to ``run gravity
backwards'' from the present day galaxy distribution. This has been done with
dynamical schemes  (eg. Peebles 1989, Nusser \& Dekel 1992,
Croft \& Gazta\~{n}aga 1997) and also by 
applying a one to one mapping between the final smoothed density and the
initial density (assumed to have a Gaussian p.d.f.)  (Weinberg 1992).
A different approach attempts to recover statistical measures of the
initial conditions (eg. the two-point correlation function or the power
spectrum) from knowledge of how gravitational evolution has affected these 
statistics. For example, formulae have been proposed which map the correlation
function measured at the present day onto the initial correlation 
function (Hamilton $\etal$ 1991). The same is true for the 
power spectrum (Peacock and Dodds 1994,  Jain, Mo \& White 1995
, Baugh \& Gazta\~{n}aga 1996).
In this paper we will show how the initial power spectrum shape  
 can be inferred from measurements of the space
density of peaks in the galaxy distribution, even on scales where significant
non-linear evolution has taken place in the underlying mass.

Peaks in the initial density distribution have been thought to be 
potential sites for the  formation of galaxies and clusters of
galaxies. Two seminal papers by Kaiser (1984) and Bardeen $\etal$ (1986) 
in which the details of the theory of Gaussian
random fields relevant to these  problems were studied have been very
influential in the study of galaxy and large-scale structure formation.
In the present paper we apply these results to the density 
field smoothed on larger scales, in order to see how peak theory describes 
the properties of an evolved density field. We are
particularly interested in seeing  whether the distribution
of peaks enables us to recover some information about the initial conditions,
and also in its potential use as a discriminator between cosmological models.

When applying peak theory to the problem of galaxy or cluster formation,
the assumption is usually made that these objects have formed from the
 gravitational collapse of matter around a high peak. Some smoothing scale,
somewhat arbitrarily chosen, is 
associated with the object to be formed. The validity of these assumptions
has been tested on the small scales relevant to the formation of individual
objects by Park (1991), and Katz $\etal$ (1993) amongst others. The
conclusions seem to be that at least for the formation of matter haloes,
there is only qualitative agreement between the predictions and the results of
numerical  simulations. Indeed in view of the extensive merging of matter 
clumps predicted by  presently favoured hierarchical models of structure
formation (see e.g. Press $\&$ Schecter 1974, Lacey $\&$ Cole 1993) it is 
reasonable to expect that evolution
of the density field on small scales will tend to disrupt the predictions
of peak theory.

 When the matter distribution is smoothed on somewhat larger
scales however, for example in the 
estimation of the genus (e.g. Melott, Weinberg $\&$ Gott 1988),
the topology and some aspects of
large scale features of the density field seem to be little
affected by gravitational evolution. 
In particular, the rank order of density contrasts measured at points in the
field is approximately conserved.
This fact has been used by Weinberg (1992) in the reconstruction method 
mentioned above. This reconstruction method also appears to work well
 when tested on simulations where the matter distribution 
is biased using some ad hoc galaxy  formation model. 

In this paper we will use numerical simulations to test whether this 
insensitivity to gravitational evolution affects the mean number
of peaks  and if
so on what scales it begins to break down. 
We are also interested in reconstructing the power spectrum shape 
from the peak density as one specifies the other entirely in Gaussian models.
The outline of the paper is as
 follows: In Section 2 we will briefly introduce the concept of the
number density of peaks in
 Gaussian random fields and in $N$-body simulations. We will describe the
simulations and our method of peak selection before testing peak theory 
against the evolved density both of mass and of `galaxies' identified in the
 models. In Section 3 (and Appendix A) 
we will describe a stable method of recovering
the power spectrum slope from the peak density. We will test it first on
analytic power spectra and then on the simulation results. In Section 4
we discuss our results and present some conclusions.

\section{The number density of peaks}
We consider local maxima of all heights in a density field to be peaks. 
In this paper we will only consider cosmological models which start from
Gaussian initial conditions. The derivation of quantities related to the 
power spectrum of fluctuations will rely on this assumption, but of course 
the raw peak density itself when measured in observations can be used to
apply constraints to other sorts of models.

\subsection{Peaks in Gaussian random fields}

For a Gaussian random field the number density of peaks per
unit volume $n_{pk}$ is given by (Bardeen \etal 1986):

\beq
n_{pk} = {29 -6 \sqrt{{6}} \over{2 (2\pi)^2 5^{3/2}}} R_*^{-3},
\label{npk}	
\eeq
where $R_*$ is the ratio of moments of the power spectrum $P(k)$
corresponding to a characterstic scale in the Gaussian field:
\beqa
R_* &\equiv& \sqrt{3} {\sigma_1/\sigma_2}, \\
\sigma_1^2 &\equiv& {1\over{2\pi^2}} \int~ dk~k^4 P(k), \nonumber \\
\sigma_2^2 &\equiv& {1\over{2\pi^2}} \int~ dk~k^6 P(k), \nonumber
\label{rstar}
\eeqa

In a practical situation the field has to be smoothed on the scale, 
$r_f$, of interest. We will use a Gaussian window filter, so that
$P(k)$ in the above equations should be multiplied
by $\exp{[-(kr_f)^2]}$.

\subsection{Peaks in numerical simulations}

To find
maxima we need the density field to be differentiable, which 
means that a filter must be applied  to the 
particle based realisation of density derived from $N$-body simulations.
An additional limitation imposed on us by simulations is their 
limited ability to resolve underdense regions, which will largely affect
 minima.
These are present in identical numbers to
maxima in Gaussian fields. The 
sampling of maxima being in general better than that of minima (most
local maxima are `higher' than most local minima) we choose to study only
maxima here.

We use simulations of three different spatially flat
 cold dark matter dominated Universes.
One set of simulations is of ``standard'' CDM, with $\Omega_{0}=1, h=0.5$,
  and another
is of low density CDM with $\Omega_{0}=0.2, h=1$ and a cosmological
constant $\Lambda=0.8 \times 3H_{0}^{2}$.
The power spectra, P(k) for these models are
 taken from Bond \& Efstathiou (1984)
 and Efstathiou, Bond and White (1992). The shape of P(k) is
parametrised by the parameter $\Gamma=\Omega h$, so that we
have $\Gamma=0.5$ CDM and  $\Gamma=0.2$ CDM.
 We also use simulations of a Mixed 
Dark Matter universe where a massive neutrino component contributes
$\Omega_{\nu}=0.3$, $\Omega_{CDM}=0.6$ and $\Omega_{baryon}=0.1$.
The power spectrum was taken from Klypin $\etal$ (1993). 
 Each simulation
contains $10^{6}$ particles in a box of comoving side-length $300 \hmpc$
and was run using a P$^{3}$M $N$-body code 
(Hockney and Eastwood 1981, Efstathiou \etal 1985). 
The mean comoving interparticle separation is therefore $3 \hmpc$, 
of the same
order as that of normal galaxies.
The simulations are descibed in more detail in Dalton \etal (1994).
In this paper we use 5 realisations of each model, error bars being
estimated from the standard deviation of their results.

\begin{figure}
\vspace{8.0cm}
\includegraphics{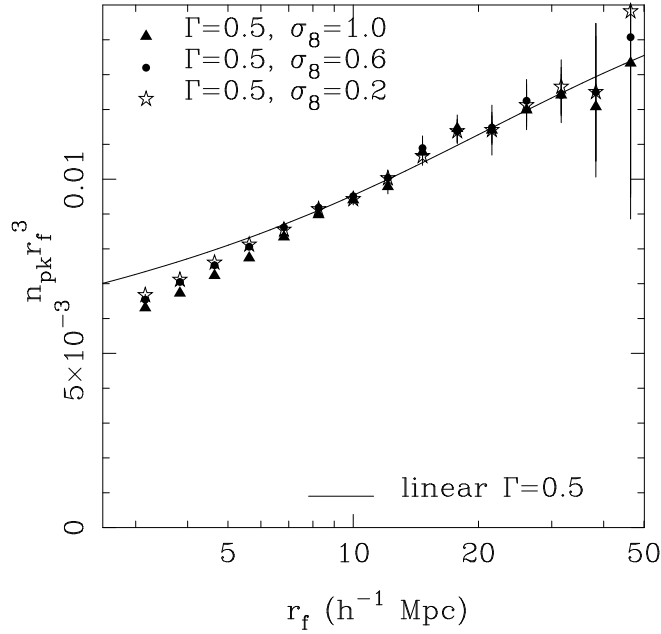}
\caption[junk]{Peak space density (actually
$n_{pk}r_{f}^{3}$, proportional to the mean number 
of peaks per smoothing volume)  for three different output times
 in the Standard ($\Gamma=0.5$) CDM scenario.
The errors on the points are calculated from the standard deviation
about the mean of results from 5 simulations.
 The linear theory
prediction is shown by a curve. 
\label{junk1} }
\end{figure}

To find peaks, we first assign the particle density to a $256^{3}$ grid using
a cloud in cell scheme ( Hockney \& Eastwood 1981). We then smooth the density 
field in Fourier space using a Gaussian filter of the appropriate radius.
We then identify local maxima on the grid in real space.
 We carry out this procedure for
 several logarithmically spaced values of the filter radius. We find that if
the filter radius is smaller than the mean interparticle separation 
the number of peaks increases, due to discreteness effects, as isolated 
particles become erroneously identified as peaks. The finite grid 
may also have a small effect in the opposite direction as we cannot identify 
peaks which are less than two grid cells apart. In this paper we choose to 
analyse fully sampled simulations, using filter radii from $3 \hmpc$ to 
$50 \hmpc$. We defer the study of discreteness effects, which will be
important in the study of real galaxy surveys, to future work.
In finding peaks, we also make use of the periodic boundary conditions
 in the simulations. The boundaries of
real surveys will also have an effect which will need to 
be quantified.

\begin{figure}
\vspace{8.0cm}
\includegraphics{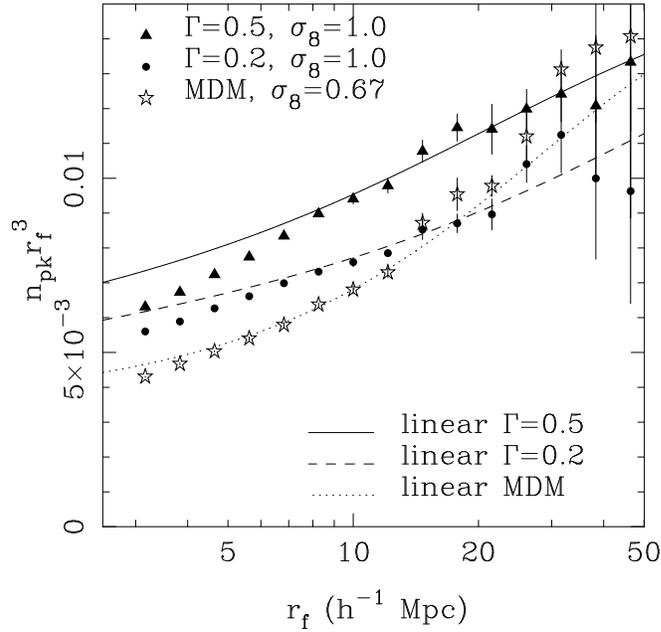}
\caption[junk]{$n_{pk}r_{f}^{3}$ for three different cosmological models,
 $\Gamma=0.5$ CDM, $\Gamma=0.2$ CDM, and Mixed Dark Matter
The linear theory predictions are shown by curves. 
\label{junk2} }
\end{figure}

Our results, the number density of peaks as a function of filter radius
are shown in Figures 1 and 2.
When plotting our results, in order to make the figures clearer,
we have chosen to plot not the peak space density  but $n_{pk}$ multiplied
by $r_{f}^{3}$. This is proportional to  the number of peaks
enclosed on average by a smoothing volume. 
 We show results for the $\Gamma=0.5$ CDM 
simulation at different stages in its evolution (Figure 1), 
as well as results for the other models (Figure 2).
The simulation points are labelled according to the predicted linear theory 
amplitude of fluctuations
in $8 \hmpc$ radius spheres, $\sigma_{8}$. The models shown in Figure 2 have a
fluctuation amplitude in the range allowed by the 4 year COBE results  
(Hinshaw \etal 1996). 
In the figures, we also show as curves
the predictions of peak  theory of Gaussian
random fields calculated using the formulae of Section 2.1.

During the course of gravitational evolution, we do expect the peaks 
to move and to 
change in height through accretion of matter. Indeed, the number 
density of objects above a certain mass threshold changes rapidly, and can be 
used as a sensitive probe of the value of $\sigma_{8}$ (see eg. White,
Efstathiou and Frenk 1993). However,  in this paper we are interested in
the total space density of peaks of all heights, which is predicted to
be constant in time in linear theory.

We can see from the first two figures that the measured number of peaks follows
the theoretical prediction remarkably well, particularily on
filter scales $r_{f} \simgt 4-6 \hmpc$.  
 We attribute to merging
the slight decrease in the number of peaks
over time which we see in Figure 1.
Of the two CDM models, the $\Gamma=
0.5$ model, which has relatively more power on small scales departs the most
from the prediction.
Merging is not a very important process, 
though, as even on scales of $3 \hmpc$, the space density of peaks only
falls $\sim 10-15 \%$ below the peak theory prediction by the time
the simulation has reached $\sigma_{8}=1$. There appears to be even less 
merging in the MDM model.  

The amplitude of fluctuations in the density field smoothed on the
scales where the peak theory prediction begins to break down is
$<\delta>\sim1-2$, which is in what is conventionally referred to as the
quasilinear regime. We are therefore probing roughly the same scales 
where  higher order perturbation theory (eg. Baugh, 
and Efstathiou 1994) can be used to estimate 
departures from linear theory.

\subsection{Biased simulations}

\begin{figure}[t]
\centering
\vspace{8.0cm}
\includegraphics{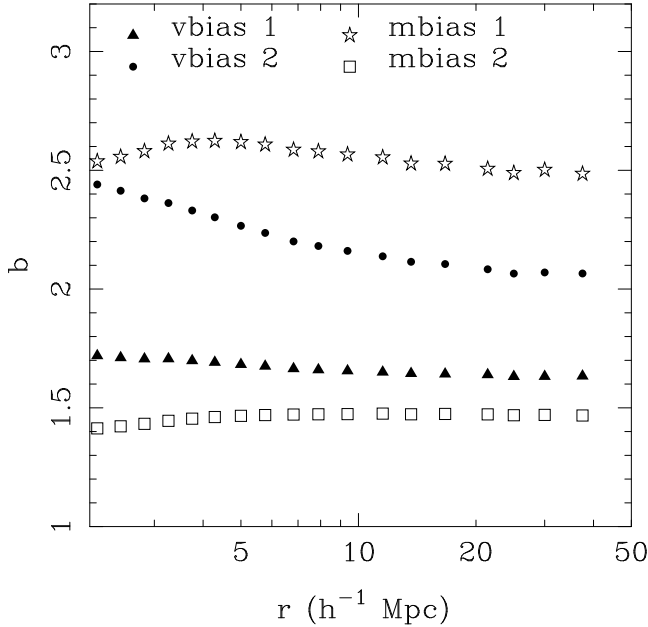}
\caption[junk]{
The effective bias $b\equiv \sigma_{galaxies}(r)/\sigma_{mass}(r)$ as a 
function of scale for 4 different biasing prescriptions applied to the
$\Gamma=0.5$ simulations (see Section 2.3).}
\label{bsc150}
\end{figure}

The relationship between galaxies and mass is one of the great uncertainties
which affects the study of large-scale structure. It would be very useful
to be able to measure properties of the galaxy density field and know
exactly how they are related to those of the underlying mass.  
In the calculation of statistical clustering measures such as
the two-point correlation function the uncertainty is parametrised with
a linear biasing parameter. The value of this parameter in the real Universe,
is still an unknown, if indeed it can be measured at all. In the case of the
peak density, we can be hopeful that differing relationships
between galaxies and  the mass will have
a smaller effect than on the two-point correlation function. It seems 
reasonable to expect that where a concentration of matter forms a peak
in the density field there will also be a peak of some sort
in the galaxy distribution. 
In this case, when we measure the total space density of peaks in the
galaxy distribution we are measuring the equivalent quantity for the
mass.
Of course it is possible to imagine that no correlation exists between 
the galaxies and mass, or that the relationship is extremely complex and
very non-local, so that our assumptions do not hold. Fortunately, there is
some evidence that this is not the case, at least on large scales.
For example, there appears to be a good correlation between galaxy
velocities  predicted from the density field and the actually 
measured velocities, which respond to the underlying mass (see eg.
da Costa $\etal$ 1997).

In order to partially test these ideas, we have applied different ad hoc
biasing prescriptions to the $\Gamma=0.5$ simulation in order to
produce a set of galaxy particles.  
One prescription involves carrying out the following procedure.
For each particle we find the distance to the 20th nearest particle and then
calculate the overdensity
inside a sphere of this radius. If the resulting fluctuation
 is above a certain
critical value (in this case 0.5) we add the particle to the list of galaxies.
This has been done for the $\Gamma=0.5$ CDM simulation, for outputs with 
linear theory mass $\sigma_{8}=0.5$ and $\sigma_{8}=0.67$. 
As the prescriptions
are based on the overdensity of a given mass of particles, we label the 
resulting results ``mbias 1'' and ``mbias 2'' respectively.
For the ``mbias 2'' prescription, we also add a random $30\%$ of 
``field'' particles to the list of galaxies, to represent  
peaks which could have formed in lower density regions 
but did not due to the finite resolution of the simulation.
The second prescription involves finding the overdensity in a fixed volume
about each particle (in this case a sphere of radius $2 \hmpc$).
We add the particle to the list of galaxies if the overdensity is
above 1.0 for one prescription or 3.0 for another. We label these sets 
``vbias 1'' and ``vbias 2'' respectively. The  $\Gamma=0.5$ CDM simulation
with $\sigma_{8}=0.5$ was used.

\begin{figure}[t]
\vspace{8.0cm}
\includegraphics{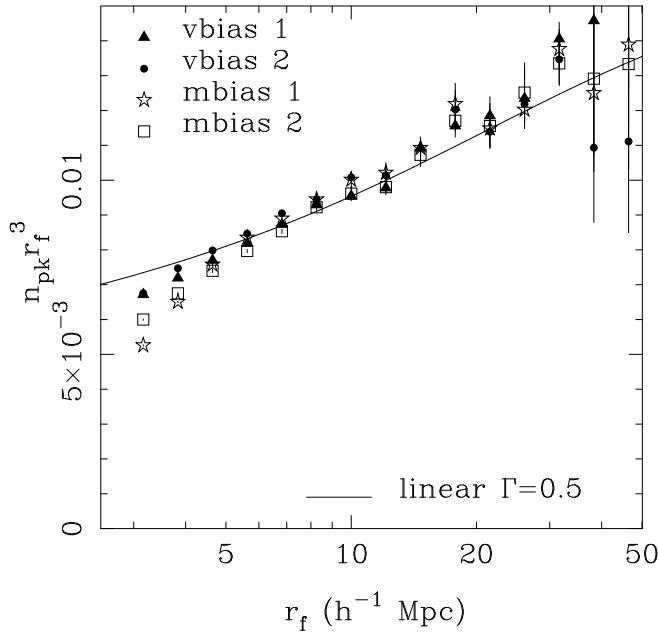}
\caption[junk]{$n_{pk}r_{r}^{3}$  for the 4 
different galaxy biasing prescriptions
(see Section 2.3 for details) applied to
 simulations of the Standard ($\Gamma=0.5$) CDM scenario. 
\label{junk3} }
\end{figure}

In all these cases, galaxy particles are preferentially selected in 
high density regions and are more clustered than the underlying mass.
In Figure (\ref{bsc150})
we show how the bias factor $\sigma(r)_{galaxies}/\sigma(r)_{mass}$
varies as a function of scale. We can see that all the models have some weak
scale dependence on small scales, with the mbias 1  galaxies having the 
largest bias factor on all scales.

The effect of these biasing prescriptions on the peak space density can be
seen in Figure 4, where we also plot the linear theory prediction for
the $\Gamma=0.5$ CDM model. For filter scales above $\sim 6 \hmpc$,
 the results 
are in good agreement with linear theory, as was the case for the unbiased 
simulations. Below this, we see a deficit in the number of peaks, which
is greatest for the mbias 1 and mbias 2 prescriptions, in which galaxies 
are selected according to the overdensity of a given number of 
surrounding particles and not governed by a physical length scale. 
For all biasing prescriptions, the deficit of peaks on 
small scales is something
 we should expect, as when we choose  galaxies,
we are effectively picking out peaks above a certain height
in the evolved mass distribution. The peaks that fall below the
threshold will disappear. The effect does not seem to be too drastic though,
and on all but the smallest scales is smaller than the differences between
cosmological models (see Figure 2).

\section{Reconstruction of the power spectrum slope}

 All the models we have tested in Section 2 started from initial conditions
with Gaussian random phases. In these cases, all statistical information
about the initial density is contained in the power spectrum, $P(k)$.
From Equations 1 and 2, we can see that the shape (but not the amplitude) of
the linear power spectrum is responsible for the space density of peaks in the
initial density field. In this Section we will present a simple and stable
method for inverting these relations and recovering the shape of  $P(k)$
from the space density of peaks. We will then apply this method to the space
density of peaks measured from simulations. Since Section 2 has shown us that
on quasilinear scales this is the same as $n_{pk}$ in the initial conditions,
we will therefore be reconstructing the initial power spectrum shape.

Given the number density
of peaks for a Gaussian field in equation (1), what can we say about 
its power spectrum?  In other words, what information about $P(k)$
can be obtained from the ratio of its moments, $R_*$, in equations 2-3?
It is clear from these equations that $R_*$ is independent of the
amplitude of $P(k)$, but it depends on its shape.
If we wanted to compare with a known
family of power spectrum shapes (e.g. CDM), we could just find the
best parameter in the family (e.g. $\Gamma$) to fit the values of
$R_*$ as a function of $r_f$.
To find the power spectrum shape more generally, one could try 
using a numerical inversion technique (eg. a generalization of the method of 
Lucy 1974) to extract the shape of $P(k)$ from the measured values of $R_{*}$.

A simpler approach, which we adopt here, is
to approximate  $P(k)$ locally with a power law $P(k) = k^n$: 

\beq
n(k) \equiv {d\log{P(k)}\over{dk}}.
\label{slope}
\eeq

If the slope $n(k)$ is constant with scale, then we substitute
$P(k)=k^n~\exp{[-(kR_f)^2]}$ into Equations 2 and 3 and  find 
(Bardeen \etal 1986):

\beq
R_* = \left({6\over{n+5}}\right)^{1/2} r_f.
\label{rstarf}
\eeq

This is only exactly true if $n$ is a constant. However, if we assume
that $P(k)$ has a slowly varying shape, then it will remain 
a good approximation when applied locally.
  $n$ is then given by the  local value
at  $k  \sim 1/r_f$,
(where the Gaussian filter gives its maximum contribution).
 In Appendix A we show how to find the best 
effective relation between $k$ and $r_f$. We can then 
 use Equation 4 to translate the estimated values 
of $R_*$ at a given $r_f$ to mean values of $n$ at a given $k$.
In Appendix A, we also test the method with various different
analytic power spectra, going from the power spectrum to the peak density
and then back to the reconstructed power spectrum slope.
We find that the method is stable and works well.

\subsection{Reconstruction from the peak density in simulations}

\begin{figure}
\vspace{8.0cm}
\includegraphics{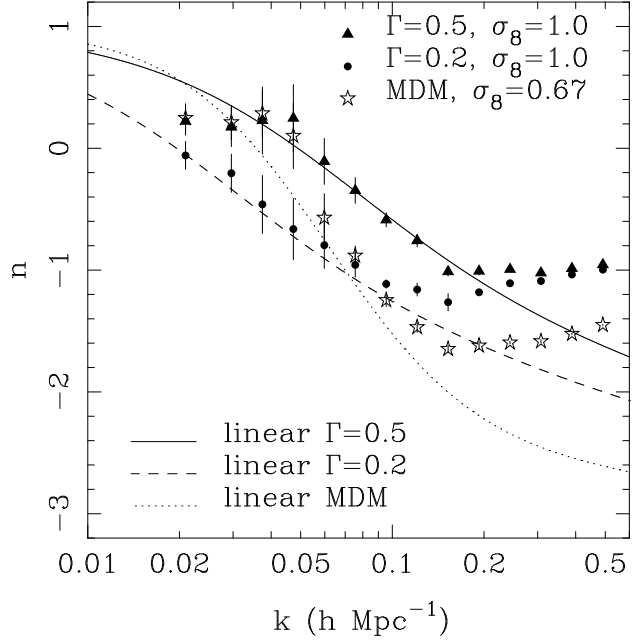}
\caption[junk]{The measured non-linear
 power spectrum logarithmic slope, $n$ for different 
cosmological models. The points are the mean of results for 
5 simulations, with the error on the mean calculated from their standard
deviation.   The curves are the linear theory values for $n$. 
\label{junk4} }
\end{figure}

When a density field undergoes non-linear evolution under gravity,
mode coupling causes a relative transfer of power between large and small 
scales. In the cases considered here, 
this has the effect of making $P(k)$ less steep on
small scales (see, eg. Baugh \& Efstathiou 1994). In Figure
9 we show $n(k)$, the logarithmic slope of $P(k)$ measured from the evolved 
mass distribution of our three different cosmological models.
This was estimated by first calculating $P(k)$ using an FFT and then 
applying a two-point finite difference operator to recover $n$ at the points
shown in the figure. We also plot the linear theory predictions for $n$,
which enables us to see that non-linear evolution of $n$ is visible
in the simulations  on scales $k \simgt 0.07 \hmpc$. From Equation A1,
using $\eta \simeq 0.4$, this scale is equivalent to $r_{f} \simlt 20 \hmpc$.

\begin{figure}
\vspace{8.0cm}
\includegraphics{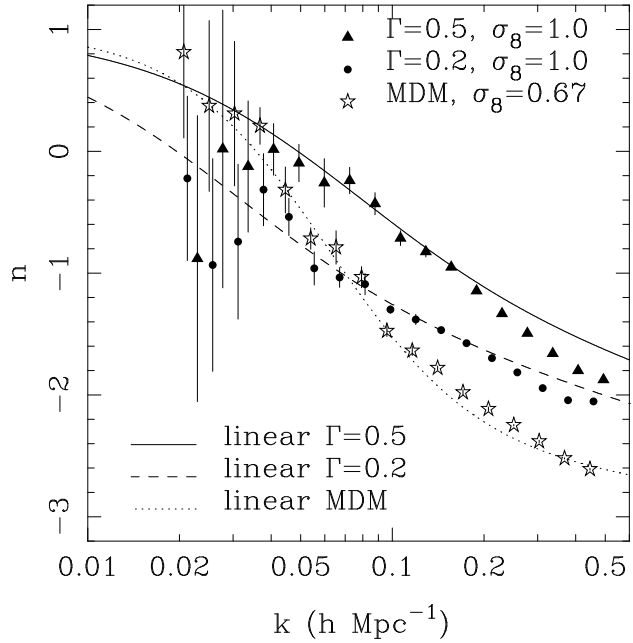}
\caption[junk]{Power spectrum logarithmic slope
, $n$  reconstructed from the peak space density 
measured from simulations of  three
different cosmological models. 
\label{junk5} }
\end{figure}

\begin{figure}
\vspace{8.0cm}
\includegraphics{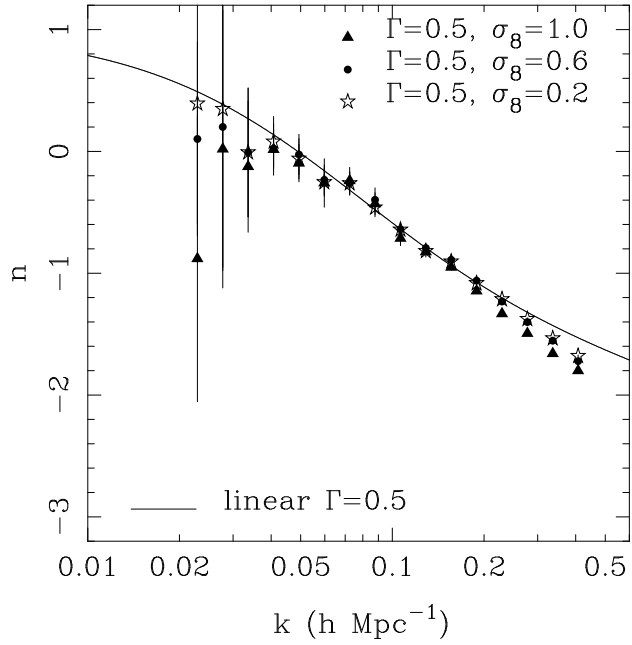}
\caption[junk]{Reconstructed power spectrum logarithmic slope, $n$ for three
 output times of the $\Gamma=0.5$ CDM simulation. 
\label{junk6} }
\end{figure}

\begin{figure}
\vspace{8.0cm}
\includegraphics{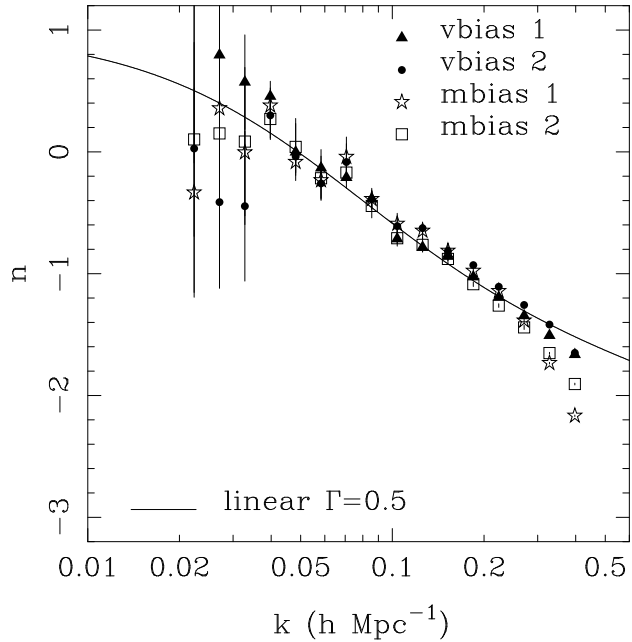}
\caption[junk]{Reconstructed power spectrum logarithmic slope, $n$ for the
different galaxy biasing prescriptions descibed in Section 2.3
and applied to the $\Gamma=0.5$ CDM simulations.
\label{junk7} }
\end{figure}

We now apply our reconstruction method described in  
 to the peak density measured
from the simulations. This results in an estimate of $n$ as a function
of scale which we plot in Figure 10. The reconstructed $n$ values are
reasonably close to the linear theory values, as we would expect.
Again, the $\Gamma=0.5$ CDM simulation suffers the most from non-linear
effects. It is interesting that the non-linearities (ie. merging) 
which affect the peak density cause it to be lowered, which means that the
recovered $n$ is slightly steeper than linear theory. 
We can see this on the smallest scales in Figure 11, in which we plot the
reconstructed $n$ for different output times in the  $\Gamma=0.5$ CDM 
simulation. This is opposite
to the effect of non-linear evolution on the directly measured value of $n$ 
(see Figure 9).  

In Figure 12 we show that the reconstruction also works well
on the peak density measured from the biased simulations, 
and again works best on larger scales.

\section{Discussion and conclusions}

Although a local transformation of a density field can change 
the  heights of peaks, it should not affect the peak number density. 
Thus, as far as the effects of both
 gravity and biasing are approximately local  
one should expect the galaxy peak density to be a preserved quantity.
Gravity is local on linear scales and there has not 
been enough cosmic time for gravity to affect (or become non-linear on)
scales much larger than $\sim 8 \mpc$. For similar reasons it also seems
difficult for biasing to be significantly non-local (see also Gazta\~{n}aga
\& Frieman 1994 and references therein).
Of course, it may  also be possible to have non-local transformations
which preserve the peak number density.

Even without attempting to reconstruct the power spectrum shape from it, the 
space density of peaks appears to be an interesting statistic. At the present
time, the two cosmological models which are the most favoured  alternatives to 
$\Gamma=0.5$ CDM are Low density CDM and MDM, which we have seen
have very differently shaped linear power spectra and different peak space 
densities. The similarity of their
shapes on quasilinear scales, and the uncertainties from biasing on
smaller scales mean that 
galaxy clustering observations generally favour
neither one or the other. We propose that observational
measurements of the peak space density would constitute a
simple means of telling them apart.

Attempts have already been made to constrain models based on linear 
power spectra reconstructed from the non-linear observations 
using mapping formulae calibrated against $N$-body simulations (as mentioned
in Section 1). Unfortunately, two different analyses have come 
up with different conclusions, each favouring a different model,
Low density CDM for  Peacock \& Dodds (1994) and
MDM for Baugh \& Gaztanaga (1996). This may be because of the sensitivity
of the  non-linear mapping formulae to cosmological parameters
such as $\Omega$, or the fact that they work less well with steep
power spectra.
The reconstruction from the peak space density should not be sensitive to the
values of $\Omega$ and $\Lambda$, for example, and should be more insensitive
to variations in galaxy biasing. It should be borne in mind, though, that
the work in this paper is aimed at the quasilinear regime and that
the non-linear mapping formulae can, in principle,
be used as a tool to study much  smaller scales (although the uncertainties
due to biasing could be large).

If one assumes that the initial conditions were Gaussian, we have shown that
it is possible to reconstruct the shape of P(k) on interesting scales
from the peak density. The method appears to work best for the models with
the steeper slopes on small scales, MDM and $\Gamma=0.2$ CDM. This is just as
well, as these models are more favoured by galaxy clustering data.
The effects of sparse sampling, boundary conditions and redshift distortions
must be studied in conjunction with an application to real data.
As with the genus statistic, a large contiguous volume is necessary
in order to estimate $n_{pk}$ reliably. With the next generation
of large redshift surveys (Colless 1997, Gunn \& Weinberg 1995)
 it should  be possible
to measure $n_{pk}$ with comparable accuracy to the simulations we have
presented here, and probably with even better accuracy on large scales.

In conclusion, we have shown that the space density of peaks in a density
field smoothed with a Gaussian filter is independent of time, following  
 the linear theory prediction, even on scales where the clustering
is mildly non-linear. We have also shown that the peak space density
in itself can be used to distinguish between cosmological models, including
the currently popular MDM and Low density CDM scenarios. The peak
density could also be used to compare with non-Gaussian models, if predictions
become available. We have developed a simple, stable method of recovering
the power spectrum from the space density of peaks and demonstrated that
the method works using analytic power spectra and the simulation results.
An application of the method to a contiguous, densely sampled galaxy survey
should yield a useful quantity, the shape of the linear power spectrum
on scales $k \simlt 0.4 \hmpc^{-1}$.

\section{Acknowledgments}
RACC would like to thank CSIC and CESCA(HCM) for financial support 
 and the members of the Institut d'Estudis Espacials
de Catalunya for their hospitality. RACC also thanks
David Weinberg for useful discussions and acknowledges support from NASA 
Astrophysical Theory Grants NAG5-2864 and NAG5-3111. 
EG thanks the Astrophysics group in Oxford
for their hospitality, and
acknowledges support from CSIC, DGICYT (Spain), project
PB93-0035 and CIRIT (Generalitat de
Catalunya), grant GR94-8001.

\setlength{\parindent}{0mm}
\vspace{3cm}
{\bf REFERENCES} 
\bigskip

\def\refe {\par \hangindent=.7cm \hangafter=1 \noindent}
\def\aj { ApJ, }
\def\mn { MNRAS, }
\def\apl { ApJ,}
\def\aa { A\&A,}

\refe Bardeen, J. M., Bond, J. R., Kaiser, N., \& Szalay, A. S. 1986, \aj,
304,15
\refe Baugh, C. M., Efstathiou, G., 1994, \mn, 270, 183
\refe Baugh, C. M., Gazta\~{n}aga, E., 1996, \mn, 280, L37
\refe Bond, J.R., Efstathiou, G., 1984, \aj, 285, L45
\refe Colless, M., 1997, in {\it Wide Field  Spectroscopy},
 eds Kontizas M., Kontizas E., Kluwer. 
\refe Croft, R.A.C., Gazta\~{n}aga, E., 1997, \mn {\it in press},
 astro-ph/9602100
\refe Dalton, G.B., Croft, R.A.C., Efstathiou, G., Sutherland, W.J.,
Maddox, S.J., Davis, M., 1994, \mn, 271, 47
\refe da Costa, L., $\etal$, 1997, in {\it Proceedings of
the 1996 Texas meeting on Relativistic Astrophysics and Cosmology}
, eds  Olinto,A.,  Frieman, J.  \& Schramm, D.N., World Scientific, Singapore.
\refe Efstathiou, G., Bond, J.R., White, S.D.M., 1992, \mn, 258, 1p
\refe Efstathiou, G., Davis, M., Frenk, C.S., White, S.D.M., 1985, 
{Astrophys. J. Suppl.}, 57, 241
\refe Gazta\~naga, E., Frieman, J.A., 1994, \apl 437, L13
\refe Gunn, J., Weinberg, D. H., 1995, in {\it Wide-Field Spectroscopy and
the Distant Universe}, eds S.J. Maddox and A. Arag\'{o}n-Salamanca,
World Scientific, Singapore.
\refe Hamilton, A.J.S., Matthews, A., Kumar, P., Lu, E., 1991, \apl, 374, L1
\refe Hinshaw, G., Banday, A.J., Bennett, C.L., Gorski, K.M.,
 Kogut, A., Smoot, G.,F., Wright, E.L., 1996, \aj 464, L17 
\refe Hockney, R. W., Eastwood, J. W., 1981, Numerical simulations
using particles (New York: McGraw-Hill)
\refe Jain, B., Mo, H.J., White, S.D.M., 1995, \mn, 276, L25
\refe Kaiser, N., 1984, \aj, 284, L9 
\refe Katz, N., Quinn, T., Gelb, J.M., 1993, \mn, 265, 689
\refe Klypin, A., Holtzman, J, Primack, J., Regos, E., 1993, \aj, 416, 1
\refe Lacey, C. G., Cole, S., 1993, \mn, 262, 627
\refe Lucy, L.B., 1974, AJ, 79, 745
\refe Melott, A. L., Weinberg, D. H., Gott, J. R., 1988, \aj, 328, 50
\refe Nusser, A., Dekel, A., 1992, \aj, 591, 443
\refe Park, C.,1991, \mn, 251, 167 
\refe Peacock, J. A., Dodds, S.J., 1994, \mn, 267, 1020p
\refe Peebles, P.J.E., 1980, {\it The Large Scale Structure of the 
Universe:} Princeton University Press
\refe Peebles, P.J.E., 1989, \apl, 344, L53
\refe Press, W.H., Schechter, P.H., 1974, 187,425
\refe Weinberg, D. H., 1992, \mn, 254, 315
\refe White, S.D.M., Efstathiou. G., Frenk, C.F. 1993, \mn, 262, 1023

\newpage

\appendix

\section{From the peak density to the power spectrum slope}

We will try a simple local reconstruction method which assumes that the
shape of $P(k)$ changes slowly with scale. In this case, we
can approximate $P(k)$ locally with a power law $P(k) = k^n$. 
  For the
case of interest, i.e. a density field smoothed with a Gaussian filter
the relation between $R_{*}$ and $r_{f}$ for a power law $P(k)$
is given by equation 4.
We have seen in Section 3 that this is only true for all $r_{f}$
if $n$ is a constant.
 If the shape is  slowly varying,
a good estimate of $n$ will be the mean local value taken assuming a one to 
one mapping between values of $k$ and $r_{f}$.
We need to find a good effective relation between $k$ and $r_f$ in order
to use equation 4 to translate the estimated values 
of $R_*$ at a given $r_f$ to mean values of $n$ at a given $k$.
The form we use is as follows,

\beq
k \equiv \eta {\pi\over{r_f}},
\eeq
where in general $\eta$ may have some dependence on the scale $k$ which will
be unimportant  if the shape of $P(k)$ varies slowly.  
For a given family of $P(k)$ the value of $\eta=\eta(k)$ can be 
found by equating the value of the local slope $n$ at $k$
with the mean value $n$ obtained from equation 4,
after finding the exact relation for $R_*=R_*(r_f)$.
In the case of a simple two power-law model with a break:

\beq
P(k) = A k^n \left[ 1 + (k_c/k)^m \right],
\label{2pw}
\eeq
we find that $\eta$ is indeed constant as a function of $k$:
\beq
\eta = {{\Gamma({5+n\over{2}})} /{{\Gamma({5+n-m\over{2}})}}} 
\label{eta2pw}
\eeq
and only depends weakly in the shape (but not the position) of the break
(at $k_c$).

\begin{figure}
\centering
\vspace{8.0cm}
\includegraphics{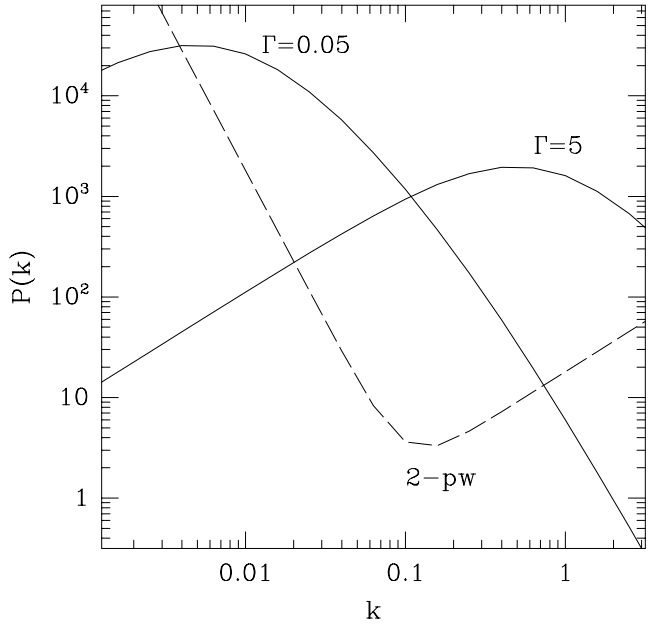}
\caption[junk]{Different shapes for $P(k)$: CDM like models with
$\Gamma=0.05$ (continuous left line) and $\Gamma=5.0$ (continuous right line),
2 power-law model (\ref{2pw}) with $n=-3$ and $m=-4$.} 
\label{pkshape}
\end{figure}

In our reconstructions we take $\eta$ to be 
a constant.
We find the best value of $\eta$ for each given $P(k)$
by finding the value which gives the smallest difference between 
the measured value of $n_{pk}$  and the value predicted
from the reconstructed $P(k)$.
We test this for a few arbitrary power spectra which we plot in 
 Figure \ref{pkshape}.
The $\Gamma$ CDM models are taken from  Efstathiou, Bond \& White (1992).

\begin{figure}
\centering
\vspace{8.0cm}
\includegraphics{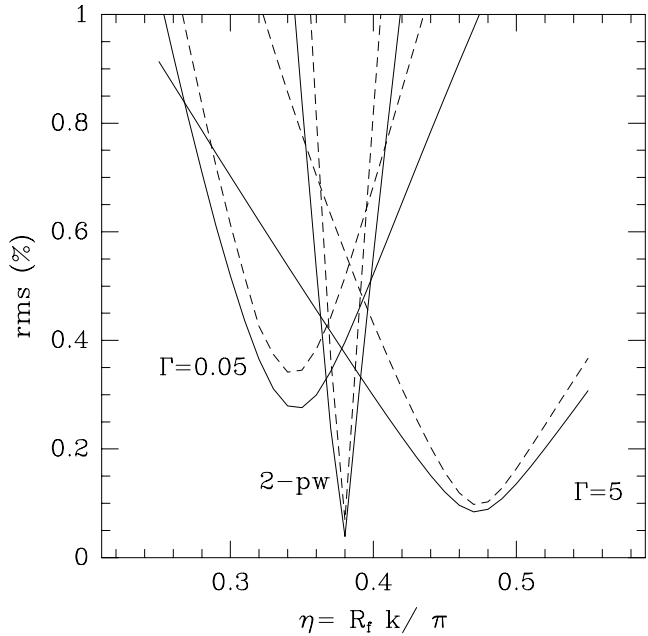}
\caption[junk]{For each of the models in Figure 7 we show
the percentage rms error per logarithmic bin in: a) $(n+5)$, where 
$n$ is the logarithmic slope of  
$P(k)$ (continuous line) b) the number density of peaks (dashed line).}
\label{eta}
\end{figure}

In Figure \ref{eta} we show, as a function of $\eta$,  
the $rms$ percentage error per logarithmic
bin in the $k$ range corresponding to $R_f=2-80 \mpc$ for: 
a) the reconstructed value of $(n+5)$ (continuous line),
compared to the input slope, and
b) the value of the number density $n_{pk}$
calculated from the reconstructed $P(k)$ 
(dashed-line), compared to the value of $n_{pk}$ calculated
directly from $P(k)$.

As expected, the two power-law model (\ref{2pw})
gives zero error at the $\eta$ predicted
by (\ref{eta2pw}). For the other models the error does not approach
zero, indicating that the true $\eta$ is not exactly a constant. Nevertheless
the minimum $rms$ error value (which defines the best value of $\eta$)
is typically small.  When applying the method to observations, as the true 
$P(k)$ is not available, we can only estimate $\eta$ by
comparing the errors between the observed and predicted 
$n_{pk}$ (dashed lines in Figure \ref{eta}), 
but as shown in the Figure this recovers well the best value of $\eta$
at the minimum $rms$.

\begin{figure}
\centering
\vspace{8.cm}
\includegraphics{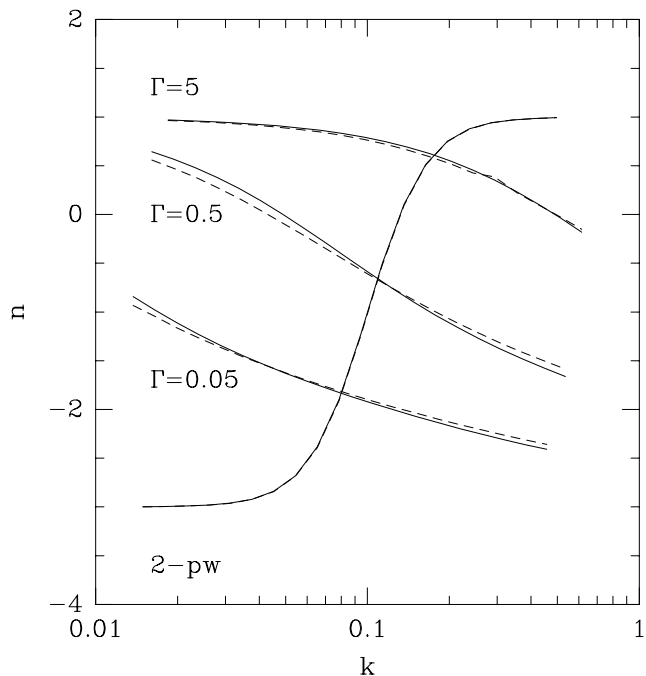}
\caption[junk]{
The reconstructed (dashed-line) and the original (continuous line) 
slope of the power spectrum for
different CDM like models with
$\Gamma=5.0$ (top), $\Gamma=0.5$ and $\Gamma=0.05$. The 
2 power-law model (\ref{2pw}) with $n=-3$ and $m=-4$ is also shown.} 
\label{nrecon}
\end{figure}

In Figure \ref{nrecon}  we show a comparison of the reconstructed
logarithmic slope (\ref{slope})
 for the values of $\eta$ that minimise the differences
between the input and reconstructed  $n_{pk}$.
The $k$ range corresponds to $R_f=2-80 \mpc$, which is
the input range in $n_{pk}$. The agreement in all cases is quite good
(less than $0.5\%  rms$  error per bin).

\begin{figure}
\centering
\vspace{8.cm}
\includegraphics{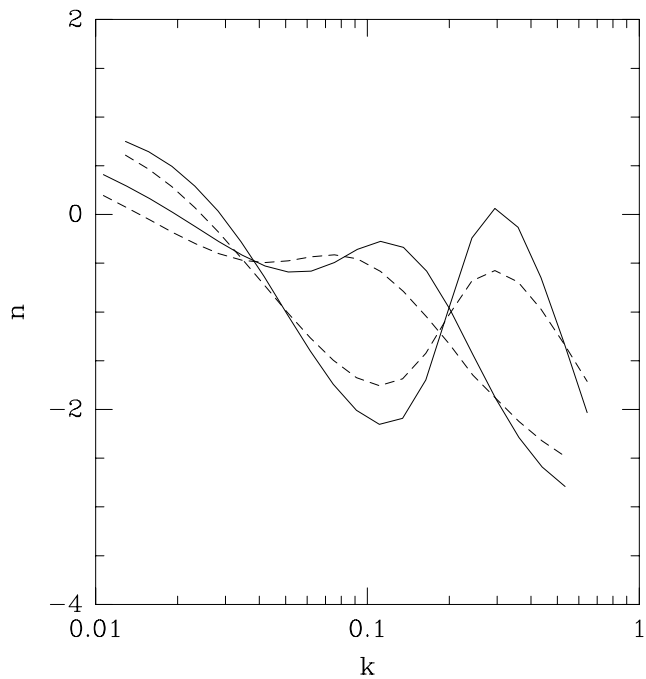}
\caption[junk]{
The reconstructed (dashed-line) and the original (continuous line) 
slope of the power spectrum for two models with rapid slope variation.}
\label{ncamel}
\end{figure}

\begin{figure}
\vspace{8.0cm}
\includegraphics{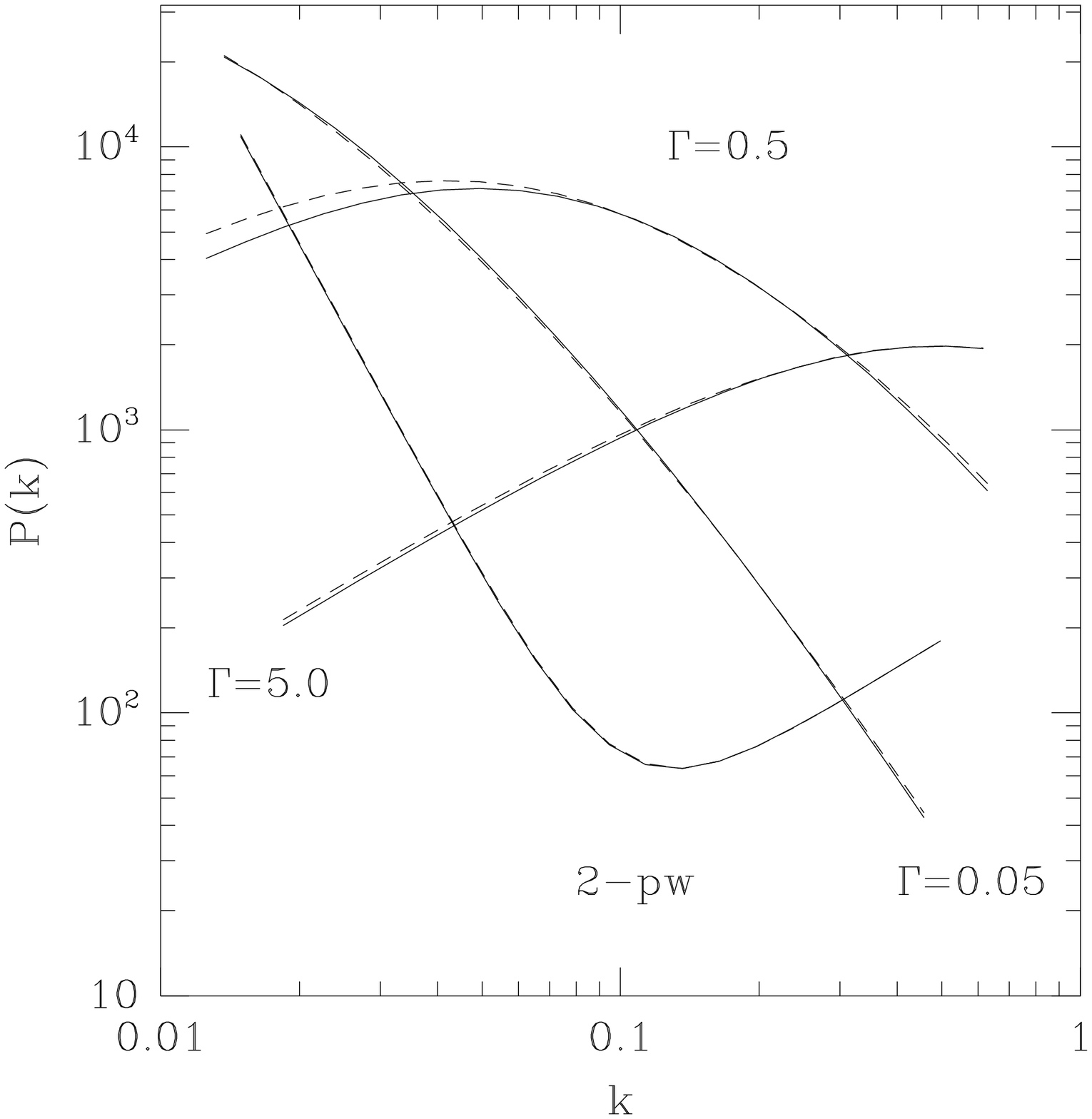}
\caption[junk]{For each of the models in Figure \ref{nrecon} we show
the reconstructed (dashed-line) and the original (continuous line) 
power spectrum.}
\label{pkrecon}
\end{figure}

In Figure \ref{ncamel} we show two ad-hoc example power spectra
 where the method
does not work quite as well as the changes in slope with scale
are more rapid.
The models shown  correspond to power spectra with 
two sharp breaks or bumps. The $rms$ error in the reconstruction is 
 larger (about  $1.5\%  rms$ per bin), but
note that the main features, such as
the positions and relative amplitudes of the breaks, 
are still recovered. In these extreme cases a more careful reconstruction
method should be used.

Once we have the shape $n=n(k)$, we can of course reconstruct
$P(k)$ itself up to an arbitrary constant by  carrying out a simple
numerical integration. This can be seen  
in Figure \ref{pkrecon} where we show a comparison between the reconstructed
and the input $P(k)$ for the values of $\eta$ that minimized the differences
between the input and reconstructed  $n_{pk}$.
To reconstruct $P(k)$ the amplitude has been matched so that $\sigma(R_f=8)=1$
with Gaussian smoothing (we have also rescaled some of the lines in
the figure for clarity). As can be seen in this figure, 
the agreement in all these cases is quite good so that $P(k)$ can
be easily recovered, up to a constant.

\end{document}